# Quantum yield optimized fluorophores for site-specific labeling and super-resolution imaging

Christian Grunwald[1*], Katrin Schulze[1*], Gregory Giannone[2], Laurent Cognet[3], Brahim Lounis[3], Daniel Choquet[2], Robert Tampé[1+]

[1]Institute of Biochemistry, Goethe-University Frankfurt, Max-von-Laue-Str. 9, D-60438 Frankfurt/M., Germany
[2] Laboratoire Physiologie Cellulaire de la Synapse, Université de Bordeaux, CNRS, 33077 Bordeaux, France
[3]Laboratoire Photonique Numérique et Nanosciences, Université de Bordeaux, Institut d'Optique Graduate School and CNRS, 33405 Talence, France

Single molecule applications, saturated pattern excitation microscopy, or stimulated emission depletion (STED) microscopy demand for bright and highly stable fluorescent dyes[1,2]. Despite of intensive research the choice of fluorophores is still very limited. Typically a stable fluorescent dyes is covalently attached to the target. This methodology brings forward a number of limitations, in particular, in case of protein labeling. First of all the fluorescent probes need to be attached selectively and site-specifically to prevent unspecific background. This often requires single cysteine mutations for covalent protein modification. Employing quantum dots allows overcoming problems of photo-bleaching[3-6]. However, the downsides are their large size, rendering the probe inaccessible to spatially confined architectures, issues in biocompatibility due to proper particle coating, and cellular toxicity[6-8]. Here we propose a new method to overcome the above outlined problems.

The reversible NTA-histidine interaction pair is widely used, but the applicability for site-specific labeling and single molecule techniques were severely hampered due to the very transient interaction ($k_{off}$ = 1 s$^{-1}$, $K_d$ = 14 µM)[9,10]. We therefore developed multivalent *bis-*, *tris-* and *tetrakis*NTA-chelators, which were rewarded by boosting the NTA-histidine affinity to the sub-nanomolar range ($K_d$ = 0.1 nM)[10]. These multivalent NTA chelator heads are modular and allow for site-specific labeling and two-dimensional organization of proteins via self-assembly[11-16].

The resolution of nanoscopic techniques is dependent on the quantum yield of the fluorescent probe[1,2]. Photo-stable fluorescent probes with high quantum yields are demanded in single molecule super-localization methods[17]. However, if the above mentioned metal-loaded chelator heads and excited fluorophores are in close proximity, the emission can be quenched by conversion into heat[13]. To maximize the quantum yield we chose polyproline-2-helices (PP2-helix) as sufficiently rigid spacer of adjustable distance[18-21]. Thereby, fluorophores and metal ions are separated by well-defined distances. Fig. 1 presents quantum yield optimized *tris*NTA-P$_x$-fluorophores with x = 0, 4, 8 and 12 prolines. Strikingly, drastically increased fluorescence signals were observed with *tris*NTA-PP2-fluorophores, which correlated with the length of the PP2-helix. In contrast, flexible OEG-spacers did not prevent the rather strong fluorescence quenching[13]. In addition to spacer length variation, different fluorophores, such as Oregon Green 488 (OG488) and ATTO565, were attached to the *tris*NTA-PP2-helices. Notably, ATTO565 is more sensitive to quenching than OG488. Quenching phenomena are generally described by the Stern-Volmer equation taking into account dynamic (e.g. by molecular collision) and static (e.g. by complex formation) mechanisms[22]. In our system the static quenching clearly dominates since the magnitude of quenching was found to be distance dependent[23]. This view is further supported by the fact that OG488 and ATTO565 showed different quenching for the same spacer length. Complementary fluorescence decay measurements were performed for different ATTO647N fluorophores, demonstrating that a P$_8$-spacer establishes fluorescence decays comparable to free and IgG-bound ATTO647N (SI Fig. 4). Omitting the prolyl spacer increased fluorescence decay rate by >40% which is in agreement with previous findings[24]. In summary, the novel *tris*NTA-PP2-fluorophores resulted in a 5-fold higher quantum yield (Table 1).

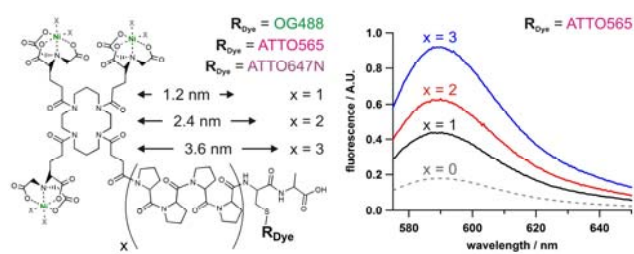

**Figure 1.** Flexible modular concept of *tris*NTA chelator heads separated from fluorescent probes (OG488, ATTO565 or ATTO647N) by rigid PP2-helices of different lengths (left). Normalized fluorescence spectra demonstrating fluorescence quenching as a function of separation distance. A fluorescence intensity of 1.0 represents an unquenched fluorophore. Results are shown for *tris*NTA-P$_x$-ATTO565 in solution (x = 0, 4, 8 and 12).

| Compound | | # of prolines | distance PP2-helix | quenching in % |
|---|---|---|---|---|
| *tris*NTA-**OG488** | (1) | 0 | 0 nm | 53 |
| *tris*NTA-**ATTO565** | (2) | | | 82 |
| *tris*NTA-P$_4$-C(**OG488**)A | (6) | 4 | 1.2 nm | 45 |
| *tris*NTA-P$_4$-C(**ATTO565**)A | (7) | | | 48 |
| *tris*NTA-P$_8$-C(**OG488**)A | (8) | 8 | 2.4 nm | 25 |
| *tris*NTA-P$_8$-C(**ATTO565**)A | (9) | | | 18 |
| *tris*NTA-P$_{12}$-C(**OG488**)A | (10) | 12 | 3.6 nm | 13 |
| *tris*NTA-P$_{12}$-C(**ATTO565**)A | (11) | | | 9 |

**Table 1.** Fluorescence quenching of *tris*NTA-P$_x$-fluorophores as a function of the distance between nickel-loaded chelator head and fluorophore. In addition, the influence of the fluorescent probe itself on the quenching phenomenon was characterized by comparing two different fluorescent dyes (**OG488** *vs.* **ATTO565**).

We next characterized the *tris*NTA-PP2-fluorophores regarding *in vivo* protein labeling and super-resolution imaging at synapses of living neurons. Here, we applied *tris*NTA-P$_8$-ATTO647N to study the distribution of AMPA receptors (GluR1) in the synapses of living neurons. These receptors within the synapse were visualized by confocal (Fig. 2) and STED microscopy (Fig. 3). As expected, the comparison of fluorescence intensity profiles reveals the superior resolution of STED microscopy. These results highlight the need for new in particular small, biocompatible, and photo-stable fluorophores in super resolution microscopy.

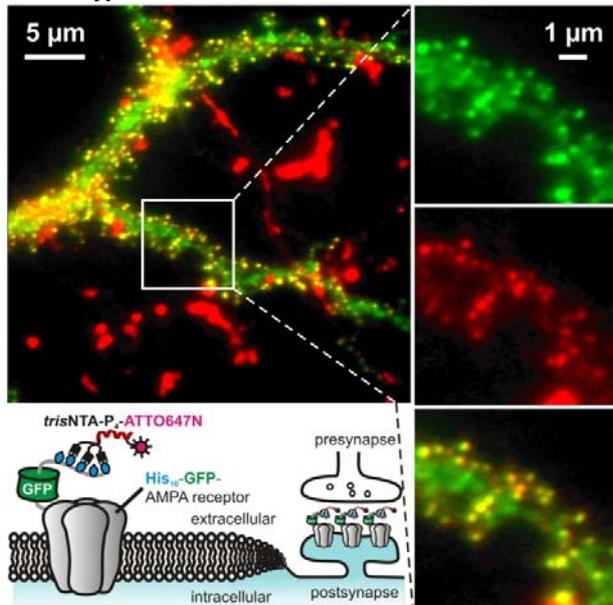

*Figure 2.* Live staining of a rat hippocampal neuron (20 d after division) transfected with $His_{10}$-GFP-GluA1 AMPA receptor subunit using trisNTA-$P_8$-ATTO647N (10 min at RT). Overlay of wide-field dual color fluorescence images. Localization of GluA1 in dendritic spines is revealed by the GFP signal (green channel; ex/em 485 ± 20 / 525 ± 40 nm) while co-localization with trisNTA-$P_8$-ATTO647N (red channel; ex/em 565 ± 30 / 655 ± 20 nm) appears in yellow. Magnified insets on the right show separated fluorescence channels demonstrating the accessibility of trisNTA-$P_8$-ATTO647N to the synaptic GluA1.

We summarize the most appealing features of our quantum yield optimized trisNTA-PP2-fluorophores: in contrast to quantum dots, no toxic effects have been reported for NTA-tagged fluorescent probes when used in vivo (e.g. intracellular particle tracking)[25]. A prevailing drawback of saturated pattern excitation microscopy is photo-bleaching of the employed fluorescent probes, which can be compensated taking advantage of the inherent reversibility of the NTA-histidine interaction. Fresh trisNTA-$P_x$-fluorophores from bulk can replenish photo-bleached fluorophores once attached to their histidine-tagged targets. A further benefit of this dynamic chemical biology approach is that imaging bandwidth is conserved even when dealing with rather fast molecular dynamics while conventional image acquisition strategies significantly struggle with maintaining good signal-to-noise levels[26]. Using quantum yield optimized trisNTA-PP2-fluorophores we were able to image and highly accurately track (<50 nm) single receptors in native cell membranes[17]. Finally, the newly developed high-quantum site-specific trisNTA-PP2-fluorophores can specifically visualize single cellular targets even in complex biological sub-structures (e.g. neuronal synapses), which are difficult to access with conventional fluorescent probes.

**Acknowledgment.** We would like to thank Gerhard Spatz-Kümbel for technical assistance. This work was funded by Centre National de la Recherche Scientifique (CNRS), the Région Aquitaine, the Agence Nationale pour la Recherche (ANR), the European Research Council (grant n 232942) to D.C. and B.L., the German Research Foundation (DFG Ta157/6 to R.T.), the BMBF (0312031/4 to R.T.) and (MODDIFSYN to D.C. and R.T.) in the frame of ERA-NET NEURON.

**Supporting Information Available:** Detailed information regarding synthesis, spectroscopic characterization, specific protein binding, and in vivo visualization of trisNTA-$P_x$-fluorophores in living neurons is given in the Supporting Information.

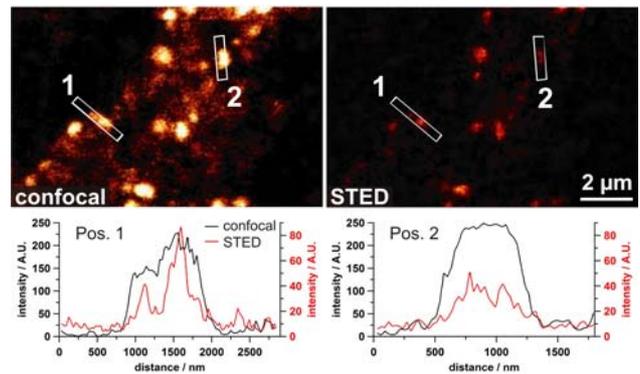

**Figure 3:** Confocal (top left) and super-resolution STED (top right) fluorescence images (ex/em 635/655 nm; depletion 750 nm) of the trisNTA-$P_8$-ATTO647N signal displaying localization of GluA1 in dendritic spines from the same section. Intensity profiles corresponding to region of interest 1 and 2 are given in the bottom panel. 100-nm synaptic clusters of GluA1 could be resolved.